\documentclass[twocolumn, pra, superscriptaddress,notitlepage]{revtex4-2}
\usepackage{graphicx}% Include figure files
\usepackage{dcolumn}% Align Table columns on decimal point
\usepackage{bm}% bold math
\usepackage[usenames,dvipsnames]{color}
\usepackage[most]{tcolorbox}
\usepackage{multirow}
\usepackage{gensymb}
\usepackage[normalem]{ulem}
\usepackage{CJK}
\usepackage{comment}
\usepackage[colorlinks, linkcolor=blue,anchorcolor=blue,citecolor=blue,urlcolor=blue]{hyperref}
\usepackage{amssymb}
\usepackage{pifont}
\DeclareUnicodeCharacter{03B8}{\ensuremath{\theta}}

\usepackage{physics}
\usepackage{natbib}
\usepackage{xcolor}
\usepackage{color,soul}
\begin{document}

\title{Nonlinear time-reversal interferometry with arbitrary quadratic collective-spin interaction}
\author{Zhiyao Hu}
\thanks{These authors contributed equally to this work.}
\affiliation{
 State Key Laboratory of Low-Dimensional Quantum Physics, Department of Physics, Tsinghua University, Beijing 100084, China
}%
\affiliation{
School of Physics, Xi’an Jiaotong University, Xi’an 710049, China
}
\author{Qixian Li}
\thanks{These authors contributed equally to this work.}
\affiliation{
 State Key Laboratory of Low-Dimensional Quantum Physics, Department of Physics, Tsinghua University, Beijing 100084, China
}%
\author{Xuanchen Zhang}
\thanks{These authors contributed equally to this work.}
\affiliation{
 State Key Laboratory of Low-Dimensional Quantum Physics, Department of Physics, Tsinghua University, Beijing 100084, China
}%
\author{He-bin Zhang}
\affiliation{
 State Key Laboratory of Low-Dimensional Quantum Physics, Department of Physics, Tsinghua University, Beijing 100084, China
}%
\author{Long-Gang Huang}
\affiliation{
 State Key Laboratory of Low-Dimensional Quantum Physics, Department of Physics, Tsinghua University, Beijing 100084, China
}%
\affiliation{China Fire and Rescue Institute, Beijing 102202, China}

\author{Yong-Chun Liu}
\email{ycliu@tsinghua.edu.cn}
\affiliation{
 State Key Laboratory of Low-Dimensional Quantum Physics, Department of Physics, Tsinghua University, Beijing 100084, China
}%

\affiliation{Frontier Science Center for Quantum Information, Beijing
100084, China}
\date{\today }

\begin{abstract}
Atomic nonlinear interferometry has wide applications in quantum metrology and quantum information science. Here we propose a nonlinear time-reversal interferometry scheme with high robustness and metrological gain based on the spin squeezing generated by arbitrary quadratic collective-spin interaction, which could be described by the Lipkin-Meshkov-Glick (LMG) model. We optimize the squeezing process, encoding process, and anti-squeezing process, finding that the two particular cases of the LMG model, one-axis twisting and two-axis twisting outperform in robustness and precision, respectively. Moreover, we propose a Floquet driving method to realize equivalent time reverse in the atomic system, which leads to high performance in precision, robustness, and operability. Our study sets a benchmark in achieving high precision and robustness in atomic nonlinear interferometry.
\end{abstract}

\maketitle

\sloppy
\preprint{APS/123-QED}

% Force line breaks with \\


%\tableofcontents

\section{Introduction}
Improving precision and system robustness is always the main task in quantum metrology and quantum sensing. Projection noise, originating from the quantum fluctuations in the measured populations, is a fundamental limitation for the improvement of precision \cite{wineland1994squeezed,pro1993quantum,pro1999pro,pro2010entanglement,liu2021limits}. To break through the limit, many-body entanglement is usually needed \cite{mb2001many,mb2008entanglement,mb2008equivalence,mb2015measuring,mb2016scalable,mb2019colloquium,mb2020entanglement,mb2017deterministic,sss2001entanglement,sss2009spin,liu2010quantum}, while spin squeezing \cite{wineland1994squeezed,kitagawa1993squeezed}  is a widely used method. Squeezed spin states (SSSs) possess good properties of reduced spin fluctuations in certain directions, thus having a variety of applications in high-precision measurements, and quantum information science \cite{sens2017quantum,sens2018beating,qi2020atom,tr2017measuring,sens2018RMP,mb2014fisher,sss2020transverse,sss2022collective,qi2021delta,sss2023squeezing,sss2023heisenberg,tr2019metrological,sss2023phase,chen2019extreme}.

Despite squeezed spin states having great properties in reducing projection noise for quantum metrology, in real experiments, the precision we could achieve is still greatly restricted by other various types of noises, including the detection noise. Under certain ranges of detection noise, the metrological gain using squeezed spin states will decrease excessively \cite{ibr2016oat}. To address this problem, an echo-like scheme similar to 
 nonlinear interferometry is proposed \cite{int19862,int2010coherent,int2012enhancement,int2015spin,int2016detecting,int2016quantum,int2017pumped,int2020quantum,int2022nonlinear,int2022time}. In this scheme,  apart from the squeezing process and encoding process, a quasi-time-reverse evolution called Interaction-Based Readout (IBR) is added before detection, which will significantly improve the robustness of the system to the detection noise \cite{ibr2016loschmidt,ibr2016oat,ibr2017optimal,ibr2018all,ibr2018TAT,ibr2020theory,ibr2022lmg,ibr2023tnt,ibr2018tnt,ibr2018cat}. The added reversal control will improve the robustness of the system to the detection noise. However, the previous studies mostly focus on IBR generated by one-axis twisting (OAT) \cite{kitagawa1993squeezed,ibr2016oat} and two-axis twisting (TAT) \cite{kitagawa1993squeezed,ibr2018TAT}, which are very special cases of collective-spin interaction. Besides, introducing the IBR scheme will lead to a tradeoff between precision and robustness, which is very important but seldom investigated. Moreover, most of the previous studies assume
that there exists a time-reversal control, which is not always easy to implement in experiments.

In this article, we investigate the performance of the nonlinear interferometry with arbitrary quadratic collective-spin interaction and propose the optimal scheme for high precision and high robustness in quantum metrology with the atomic system. To improve the precision and robustness of the system, we find the optimal squeezing process, phase encoding process, and anti-squeezing process with different anisotropic parameters for quadratic interaction. Through both analytical and numerical analysis, we also demonstrate the highest precision and highest robustness could be achieved simultaneously for the TAT interaction. Moreover, we propose a Floquet driving method to achieve the equivalent time-reversal control, which is implementable for arbitrary quadratic collective-spin interaction. We also show
that the method is robust to imperfect pulse from the Floquet driving according to numerical simulations.

The paper is organized as follows. In Sec.~(\ref{sec:2}) we first explain how the nonlinear interferometry works, introducing our system model and optimizing the interferometer processes with quadratic collective-spin interaction, which can be described by the LMG model.
In Sec.~(\ref{sec:3}), we propose that we could generate equivalent time reversal by making use of Floquet driving. In Sec.~(\ref{sec:4}), we
show that the method is robust to different kinds of noise, followed by the summary.

\begin{figure*}[htbp]
\includegraphics[width=\textwidth]{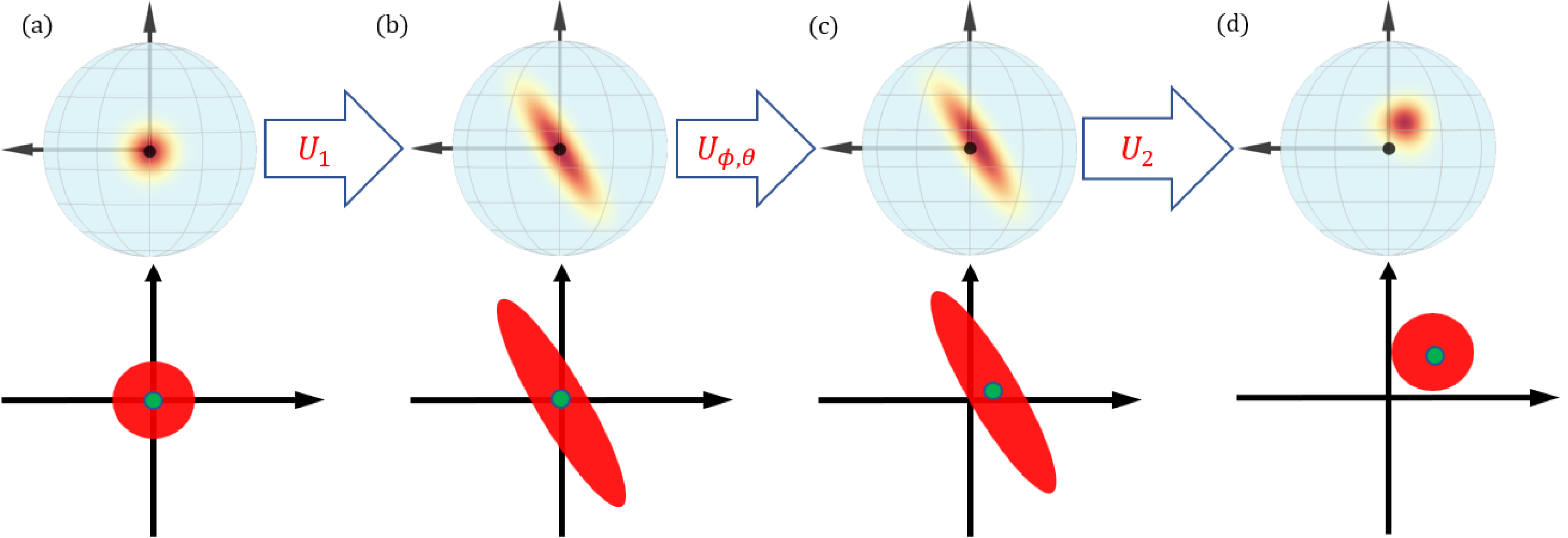}
\caption{\label{Fig:1}  The whole process of the (nonlinear) interferometry could be viewed as three steps. ({\romannumeral1}) A probe state (CSS) is prepared and it goes through the squeezing process $U_1$. ({\romannumeral2}) After the squeezing process $U_1$, an unknown parameter $\phi$ is encoded into the state through $U_{\phi,\theta}=e^{-i\phi S_{\theta}}$ along $S_{\theta}$, resulting in a small displacement in the Bloch Sphere. ({\romannumeral3}) Then an anti-squeezing process $U_2$ is performed, which is immediately followed by the measurement. The schematic diagram of this process is shown in the bottom row.
}
\end{figure*}

\section{\label{sec:2}System Model and Nonlinear Interferometer Scheme}

As shown in Fig.\ref{Fig:1}, the whole process of the nonlinear interferometer scheme could be viewed as three parts: First, a coherent spin state (CSS) is prepared and it evolves under $U_1$ for $t_1$, in which the entanglement is generated and the state turns to squeezed or oversqueezed state. Following that is the encoding part, where the estimated value $\phi$ is encoded through $U_{\phi,\theta}$. After that comes the readout part, during which the squeezed spin state evolves under a time-reversed dynamics $U_2$ for $t_2$, followed by a measurement. Now we consider how to optimize this scheme to achieve high precision and high robustness.

\subsection{\label{sec:level2}System model}
Consider a system of mutually interacting spin-1/2 particles described by
the following Hamiltonian:
\begin{equation}
H = \sum_{j,k} \chi_{\alpha\beta}\sigma_{\alpha}^{j}\sigma_{\beta}^{k},
\label{eq:one}
\end{equation}
where $\sigma_{\alpha}^{j}$ is the Pauli operator of the $j$-th spin and $%
\alpha,\beta\in\{x,y,z\}$. The parameter $\chi_{\alpha\beta}$ characterizes
the strength of the interaction in different directions. To ensure the
Hermicity of the Hamiltonian, we have $\chi_{\alpha\beta} =
\chi_{\beta\alpha}$. Here we have the assumption that the interactions between individual spins are the same, and this assumption holds when there are all-to-all interactions, which is valid under some systems such as nuclear system \cite{LMG1965validity}, cavity QED \cite{int2022time}, ion trap \cite{bohnet2016quantum}.

Now we introduce the collective spin operators $S_{\alpha} = \frac{\hbar}{2}
\sum_{j}\sigma_{\alpha}^{j}$ with commutation relations $[{S}_{\alpha},  S_{\beta}] = i\hbar \epsilon_{\alpha \beta \gamma} {S}_{\gamma},$ where $\alpha ,\beta ,\gamma$ denote the components in any three orthogonal directions, and $\epsilon_{\alpha \beta \gamma}$ is the Levi-Civita symbol, and we let $\hbar=1$. $H$ preserves the
magnitude of the total spin $S^{2} = \sum_{\alpha}S^{2}_{\alpha} $, namely, $[H,S^2] = 0$.
By applying a linear
transformation to the collective spin operators and redefining the coordinate axes we can prove that this Hamiltonian is equivalent to 
\begin{equation}
H_{\mathrm{LMG}}(\chi, \gamma) = \chi (S_{x}^2 + \gamma S_{y}^2),  0\le\gamma \le0.5
\end{equation}
which could be described as the Lipkin-Meshkov-Glick (LMG) model \cite{LMG1965validity,hu2023spin}. In the expression of $H_{\mathrm{LMG}}(\chi, \gamma)$, $\chi$ is the strength of the interaction which describes the rate at which the system evolved, and $\gamma$ is the anisotropic parameter reflecting the symmetry of the system. OAT and TAT could be viewed as special cases when $\gamma=0$ and $\gamma=0.5$ ~\cite{TAT}.

\subsection{\label{sec:level2}Squeezing process}

The first step of the nonlinear interferometer scheme is to generate entanglement. Consider a system with N interacting spin-1/2
particles, we choose a CSS as
the initial state, which can be described by
\begin{equation}
\ket{\vartheta,\varphi} = e^{i\vartheta(S_{x}\sin\varphi - S_{y}\cos\varphi)}\ket{j,j},
\end{equation}
where $\vartheta$ is the angle between the $z$-axis and the collective-spin
vector (polar angle), while $\varphi$ is the angle between the $x$-axis and the
vertical plane containing the collective-spin vector (azimuth angle). In the following, we set $\vartheta = \pi/2$ and $\varphi = \pi/2$, namely $\left | \hat{\boldsymbol {y}}\right \rangle$, as the initial state.

Firstly, the system undergoes a unitary evolution $U_{1}$, namely dynamics with the LMG Hamiltonian $H_{\mathrm{LMG}}(\chi, \gamma)$ for $t_{1}$ in our strategy, evolving to a squeezed or oversqueezed state. For the system with different anisotropic parameters $\gamma$, the corresponding $t_{1}$ to get the state with the highest precision varies as well. We need to optimize the evolution time $t_{1}$ in order to get the highest precision for estimating the value of the unknown parameter, which is bounded by quantum Cramer-Rao bound (QCRB) \cite{qfi1994statistical}:
\begin{equation}
(\Delta \phi )^2 \geq \frac{1}{kF[\rho_{\lambda}]},
\label{eq:two}
\end{equation}
where $k$ is the number of independent repetitions (in the following we set $k=1$), and $F[\rho_{\lambda}]$ is the quantum Fisher information (QFI) \cite{qfi1925theory,qfi1996generalized,qfi2006quantum,qfi2011advances} defined as:

\begin{figure}[tb]
\centering
\includegraphics[width = 0.45\textwidth]{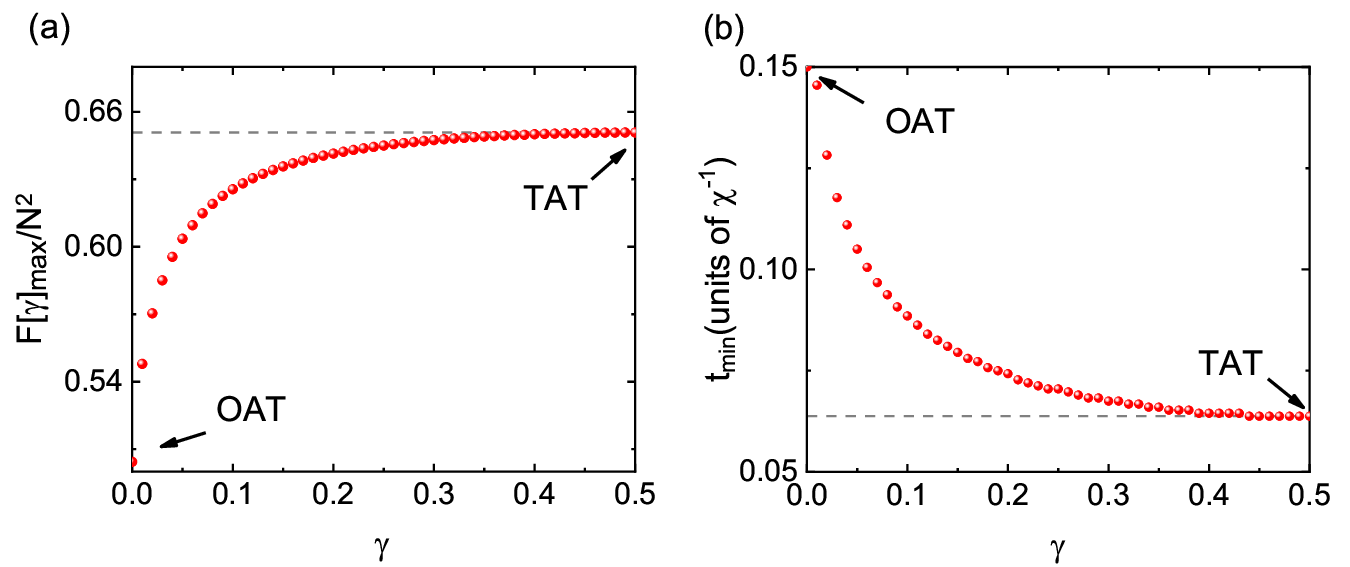}
\caption{(a)  Maximum QFI for different
anisotropic parameter $\gamma$ with $t_1$ around the best squeezing time. (b) The corresponding $t_{1}$ to get $F[\gamma]_{\mathrm{max}}$ for different $\gamma$. The horizontal dashed lines correspond to the results of the TAT model (we choose the atom number N = 100). }
\label{fig:2}
\end{figure}

\begin{equation}
F[\rho_{\lambda}] = \sum_{\kappa,\kappa^{'}}\sum_{q_{\kappa}+q_{\kappa^{'}}>0}
\frac{2}{q_{\kappa}+q_{\kappa^{'}}}
\left |  \langle \kappa^{'} | \partial_{\lambda}\rho_{\lambda}  | \kappa \rangle   \right|^2 ,
\end{equation} 
where $\hat{\rho_{\lambda}} = \sum_{\kappa} q_{\kappa} \left  | \kappa\right \rangle \left  \langle \kappa\right |$ is the $\lambda$-dependent density matrix and both the eigenvalues $q_{\kappa} \geq 0$ and the associated eigenvectors $\left  | \kappa\right \rangle$ depend on $\lambda$, and $\lambda$ is the known parameter to be estimated. For the system and parameterization we are interested in, as a pure state, the QFI simply becomes:  

\begin{equation}
F[S_{\vec{n}}] = 4(\Delta S_{\vec{n}})^2,
\end{equation} 
where $S_{\vec{n}} = \sin\vartheta \cos\varphi S_{x} +  \sin\vartheta \sin\varphi S_{y} + \cos\varphi S_{z}$. To set a higher bound for the precision of parameter estimation, $t_1$ should be set as the time realizing the highest QFI. For a certain $\gamma$, change the evolution time $t^{\prime}_{1}$ and find the corresponding maximum QFI: $F[t^{\prime}_{1}, \gamma]$. The maximum of $F[t^{\prime}_{1}, \gamma]$ then is the largest QFI with this $\gamma$, denoted as $F[\gamma]_{\mathrm{max}}$, and the corresponding $t^{\prime}_{1}$ is the $t_{1}$. Now we change the $\gamma$, and track how $F[\gamma]_{\mathrm{max}}$ and $t_{1}$ changes when
$\gamma$ varies from 0 to 0.5.  

The results are shown in Fig.~\ref{fig:2}. We find that $F[\gamma]_{\mathrm{max}}$ and $t_{1}$ monotonically depend on $\gamma$ for $0 \leq \gamma \leq 0.5$. When $\gamma = 0.5$,
the system attains its maximum $F[\gamma]_{\mathrm{max}}$, and the corresponding $t_{1}$ is also the shortest, corresponding to the
TAT interaction. What is worth mentioning is that the range of $t_1$ we choose is around the best squeezing time, while the OAT interaction $H_{\rm OAT}=\chi S_x^2$ could generate GHZ states at $t=\pi/2\chi$, the QFI of which could reach Heisenberg limit.

\begin{figure}[tb]
\centering
\includegraphics[width = 0.45\textwidth]{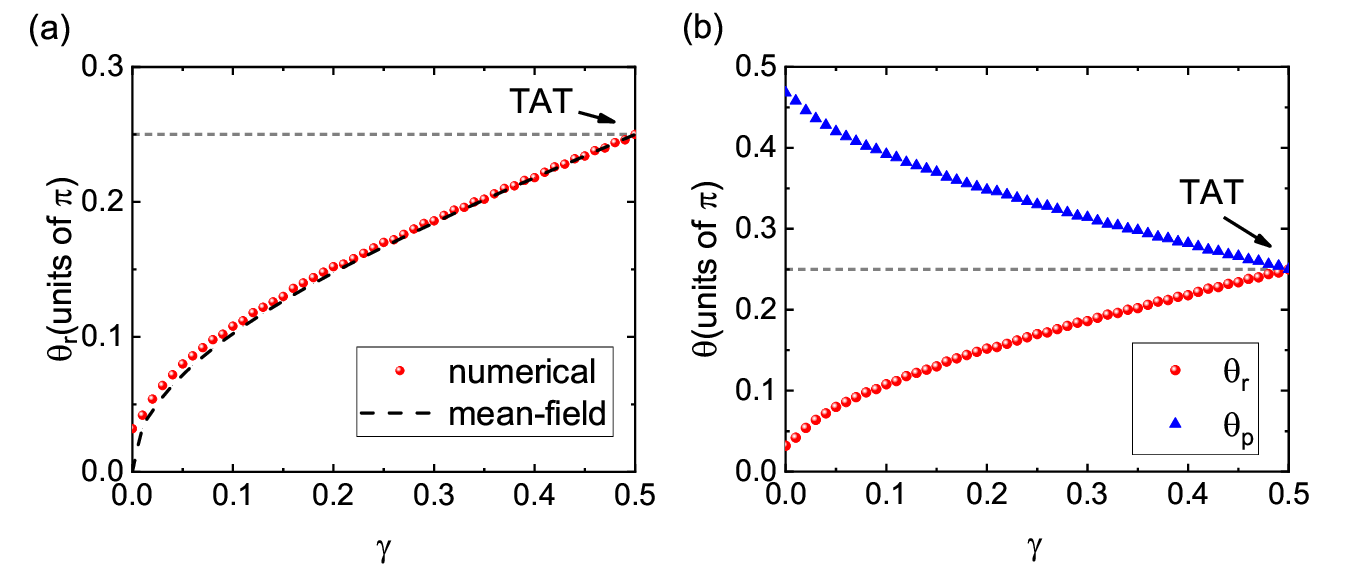}
\caption{(a)  Numerical (red circle) and analytical results through mean-field approximation (black dashed line) of $\theta_{r}$ for different $\gamma$. (b) $\theta_{r}$ (red circle) and $\theta_{p}$ (blue triangle) for different $\gamma$. The horizontal short dashed lines correspond to the results of the TAT interaction (N = 100). }
\label{fig:3}
\end{figure}
\subsection{\label{sec:level2}Encoding process}

The entanglement of the system is generated through $U_{1}$, followed by the application of the perturbation spin rotation $U_{\phi,\theta} = e^{-i\phi S_{\theta}}$, which also plays an important role in the nonlinear interferometry. Contrary to the traditional belief that the encoding part only has an effect on the precision so that the axis chosen to encode the parameter could simply choose the direction with minimum fluctuation, the encoding part has a vital influence on the robustness of the system to the detection noise if an anti-squeezing operator $U_2$ is applied before performing the measurement. 

In our model, the squeezing happens in the $x$-$z$ plane, so $S_\theta$ should be chosen in this plane to fully exploit the metrology-enhanced property. $S_{\theta}$ could be expressed as $S_{\theta} = S_{x}\sin\theta  + S_{z}\cos\theta $, where $\theta$ is the angle between the $z$-axis and $S_{\theta}$. The sensitivity for estimating $\phi$ and the robustness to detection noise of the system vary as $\theta$ takes different values. In the actual experiments, different measurements have different emphases. In some conditions, the precision might be the priority, while under certain circumstances the robustness against the measurement noise deserves extra attention.  To quantify the robustness of the system to the detection noise, here we introduce the magnification factor $G$ as:
\begin{equation}
    G=\frac{\langle S^{\phi}_m \rangle}{\frac{N}{2} \cdot \phi}.
\end{equation}
the numerator (denominator) represents the magnitude of the mean-spin measurement with (without) the time-reversal control, $S_m$ is the spin component that we apply collective mean-spin measurement and $m$ is a continuous parameter corresponding to an angle in the $x$-$z$ plane, and $G$ represents the phase magnification benefited from the time-reversed dynamics (here we assume that a perfect time-reversed dynamics is followed after the encoding process). A higher $G$ indicates the influence of detection noise will have less influence on the signal, indicating the robustness to detection noise \cite{ibr2016oat,ibr2018TAT}. 

We denote the corresponding $\theta$ as $\theta_{r}$ and $\theta_{p}$ with the directions that achieve the highest precision for estimation and robustness to detection noise, respectively. As $\gamma$ changes, $\theta_{r}$ and $\theta_{p}$ would change as well. The analytical solution of the dependence of $\theta_{r}$ and $\theta_{p}$ with respect to $\gamma$ is too complex to conduct, but we can give a brief and reasonably accurate estimate of $\theta_{r}$. We use mean-field approximation to describe the time evolution after the parameter is encoded. The detail is shown in Appendix~\ref{app:app1}. The calculation results in a brief expression of $\theta_{r}$:
\begin{equation}
\theta_{r} = \arcsin\sqrt{\gamma},
\end{equation}
which has been proved to have enough accuracy when $\gamma$ is not too small. The comparison of the numerical solution and mean-field results is shown in Fig.~\ref{fig:3}(a). We can see that our approximation is accurate enough, especially when $\gamma$ approaches 0.5. When $\gamma = 0.5$, $\theta_{r}$ becomes $\pi/4$, which is consistent with previous research \cite{ibr2018TAT}. 

\begin{figure}[tb]
\centering
\includegraphics[width = 0.4\textwidth]{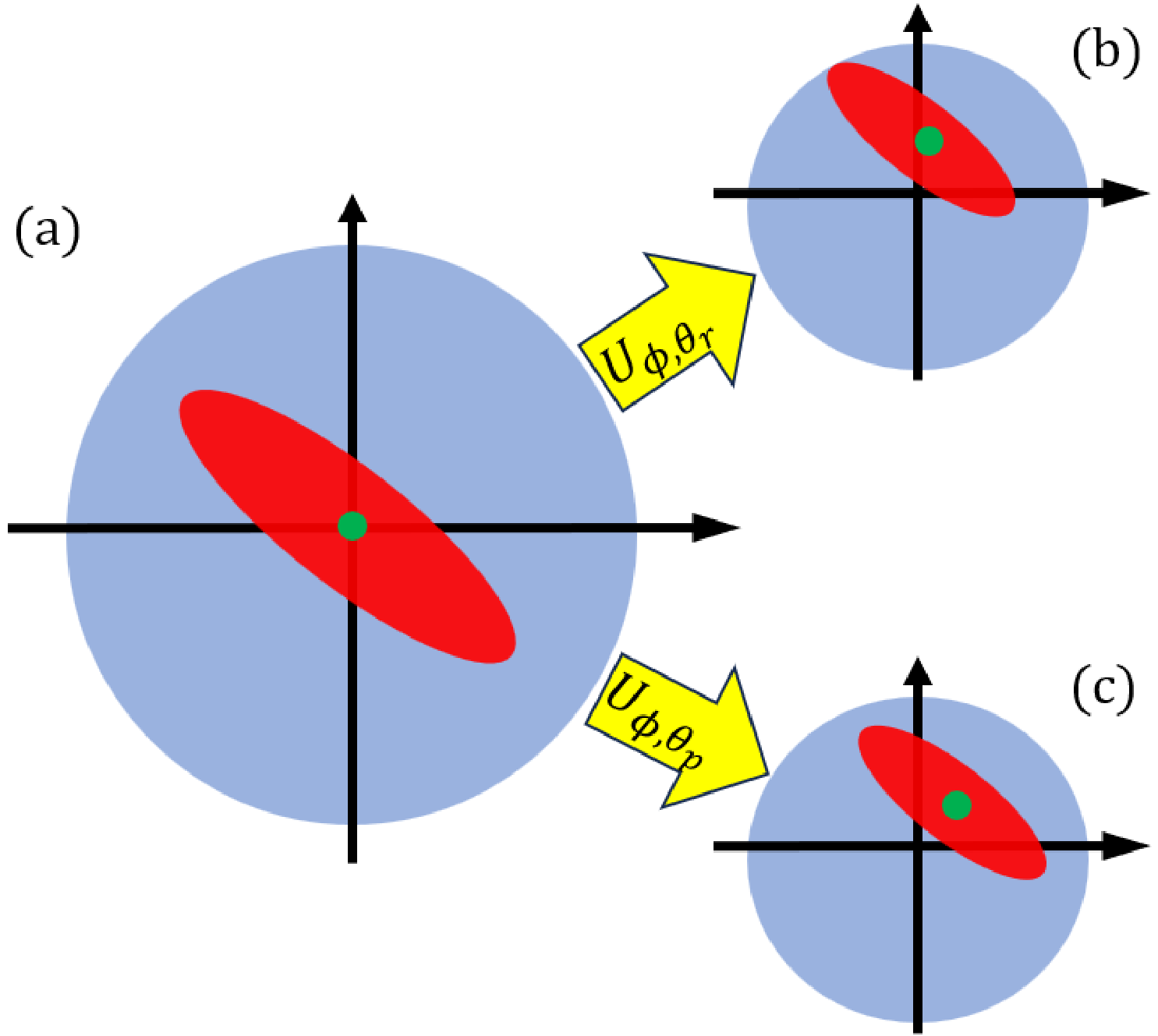}
\caption{After applying $U_1$ to the CSS, we get an SSS as the probe state. To encode the unknown parameter $\phi$, $U_{\phi,\theta}=e^{-i\phi S_{\theta}}$ is applied to the SSS, and $U_{\phi,\theta_r}$ ($U_{\phi,\theta_p}$)  is chosen to encode the parameter, which maximizes the robustness (precision) of the process. (a) the SSS generated through $U_1$. (b) the SSS after $U_{\phi,\theta_{r}}$. (c) the SSS after $U_{\phi,\theta_{p}}$. }
\label{fig4}
\end{figure}

To get the estimated value of $\theta_{p}$, we perform the numerical analysis by calculating the fluctuation for each $\theta$ after $U_{1}$ and finding the minimum one, taking the corresponding $\theta$ as $\theta_{p}$ for this $\gamma$. The result is shown in Fig.~\ref{fig:3}(b). Compared to $\theta_{r}$, we show that for a certain $\gamma$, the corresponding $\theta_{r}$ and $\theta_{p}$ have an approximate relationship:
\begin{equation}
\theta_{r} + \theta_{p} \approx \pi/2. 
\end{equation}
We find that $\gamma = 0.5$ is a special situation when $\theta_{r}$ and $\theta_{p}$ take the same value $\pi/4$, which indicates that for the TAT interaction, we can achieve the best robustness and precision simultaneously by setting $S_{\theta} = \frac{\sqrt{2}}{2} S_{x} + \frac{\sqrt{2}}{2} S_{z}$. 

\subsection{\label{sec:level2}Anti-squeezing process}

The time-reversed nonlinear dynamics is the key to our strategy. Applying $H_{\mathrm{LMG}}(-\chi, \gamma)$ for $t_{2}$,  we could perform the anti-squeezing process to get the final state to measure. An appropriate $t_{2}$ is important in this step. To determine the optimal $t_2$ to achieve higher precision, we define the metrological gain as $\Delta G$ to investigate the effect of $U_{2}$ for $t_{2}$ \cite{ibr2023tnt}:
\begin{equation}
\Delta G = -20\lg\{ \frac{\Delta \phi}{(\Delta \phi)_{\rm SQL}}\},
\end{equation}
where $(\Delta \phi)_{\mathrm{SQL}} = 1/\sqrt{N}$ is the standard quantum limit (SQL) \cite{sql2004quantum} and $\Delta \phi$ is given by error propagation formula:
\begin{equation}
\Delta \phi =  \left[\frac{\Delta S_{m}^{\phi}}{\partial_{\phi} \langle S_{m}^{\phi} \rangle }  \right] _{\phi = 0},\label{14}
\end{equation}
where $\Delta S_{m}^{\phi}$ is the standard
deviation of $S_{m}$ and $\partial_{\phi} = d/d\phi$. $\partial_{\phi} \langle S_{m}^{\phi} \rangle$ can be replaced by its first order approximation with respect to $\phi$:
\begin{align}
    \langle S_{m}^{\phi} \rangle & = \left  \langle \boldsymbol {y}\right | U_{1}^{\dagger} U_{\phi,\theta}^{\dagger} U_{2}^{\dagger} S_{m} U_{2} U_{\phi,\theta} U_{1} \left  | \boldsymbol {y}\right \rangle \nonumber \\
    & = \left  \langle \boldsymbol {y}\right | U_{1}^{\dagger} e^{iS_{\theta}\phi} U_{2}^{\dagger} S_{m} U_{2} e^{-iS_{\theta}\phi} U_{1} \left  | \boldsymbol {y}\right \rangle \nonumber \\
    & = i\phi \left  \langle \boldsymbol {y}\right | [S_{\theta}(U_{1}),S_{m}(U_{2}U_{1})] \left  | \boldsymbol {y}\right \rangle + O(\phi^2),\label{15}
\end{align}
where we define $S(U) = U^{\dagger}SU$ for brevity. 

\begin{figure}[tb]
\centering
\includegraphics[width = 0.45\textwidth]{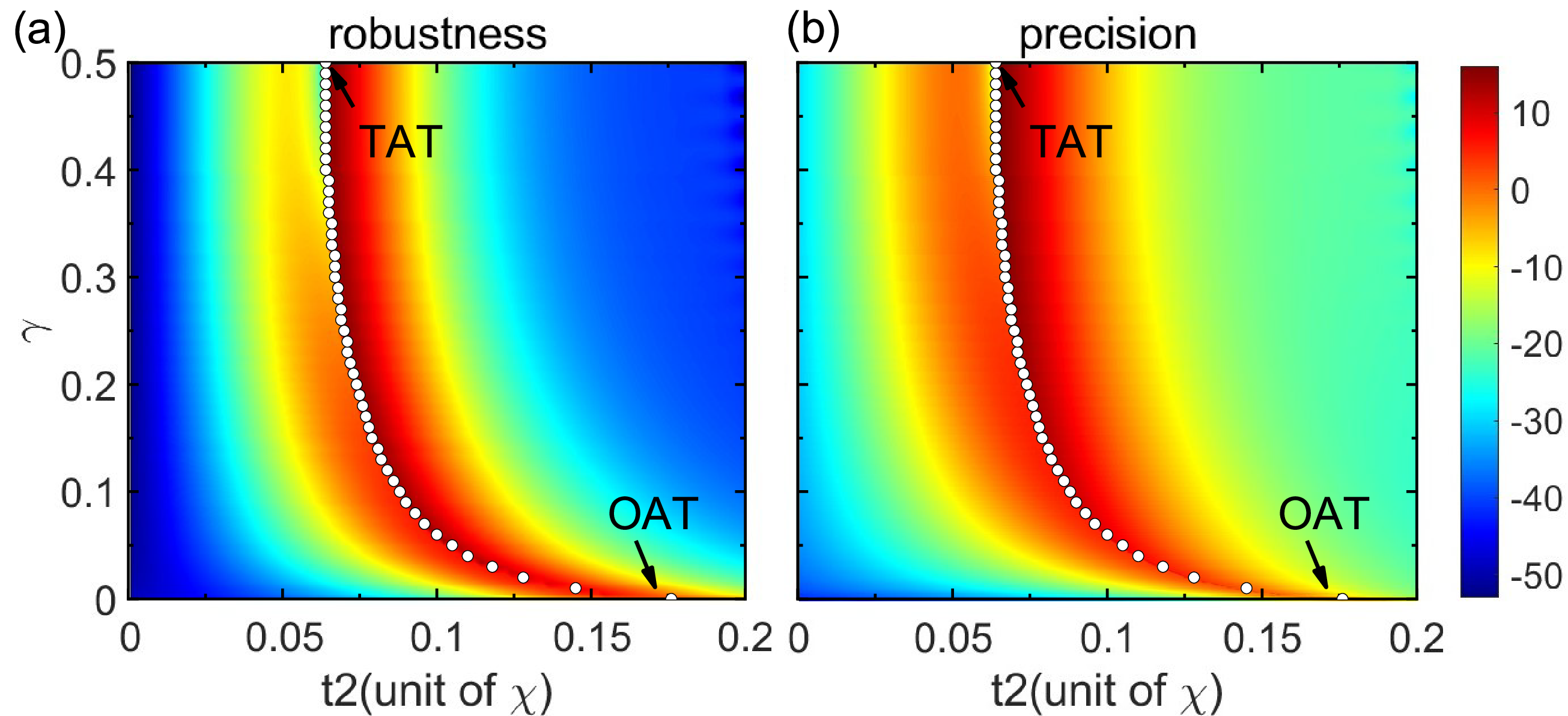}
\caption{
$\Delta G$ as a function of $t_2$, $\gamma$ and $\theta$, and here $\theta$ is chosen to optimize (a) robustness. (b) precision. The white dots represent optimal $t_1$ with $ 0\leq\gamma\leq0.5$.}
\label{fig:5}
\end{figure}

Previously we proposed two directions of $S_{\theta}$, and the dependence of metrological gain on $t_{2}$ and $\gamma$ with different directions of $S_{\theta}$ is shown in Fig.~\ref{fig:5}, respectively. We show that with $\gamma$ ranging from 0 to 0.5, to reach the maximum metrological gain, the time $t_2$ could always be chosen near $t_{2} = t_{1}$, which is conducive for us to determine the reverse time $t_{2}$ for convenience. Besides, we find that TAT outperforms OAT in precision with the reversed dynamics when the unknown parameter is encoded through the axis of $S_{\theta_{r}}$ and $S_{\theta_{p}}$, which is consistent with the previous study \cite{ibr2018TAT}.

\section{\label{sec:3}Generation of equivalent time reverse}

Realizing the interaction-based readout scheme requires implementable time-reversal control, which is usually not a trivial task in experiments. Inspired by the previous study, we propose that we could realize equivalent time reversal by making use of multiple ${\pi}$/2 pulses, which can be realized using the coupling term $\Omega_{\alpha}S_{\alpha}$ ($\alpha=x,y,z$) \cite{liu2011tat,hu2023spin,zhang2023fast}. By making use of a multi-pulse sequence along the $\alpha$-axis
($\alpha=x,y,z$), we can rotate the spin along the $\alpha$-axis and affect
the dynamic of squeezing. A $\pi/2$ pulse corresponds to $%
\int_{-\infty}^{+\infty} \Omega_{\alpha}(t)dt=\pi/2 $, which leads to the
result that $R_{\alpha,-\pi/2}e^{it\chi S^{2}_{\beta}}R_{\alpha,\pi/2}=e^{it\chi S^{2}_{\kappa}}$, where $R_{\alpha,\theta}=e^{-i\theta S_{\alpha}}$ and $\kappa$ is
the axis that perpendicular to the $\alpha$-axis and $\beta$-axis. The multi-pulse sequence is periodic, and the frequency is determined by $\gamma$ and the axis we choose.

\begin{figure}[tb]
\centering
\includegraphics[width = 0.45\textwidth]{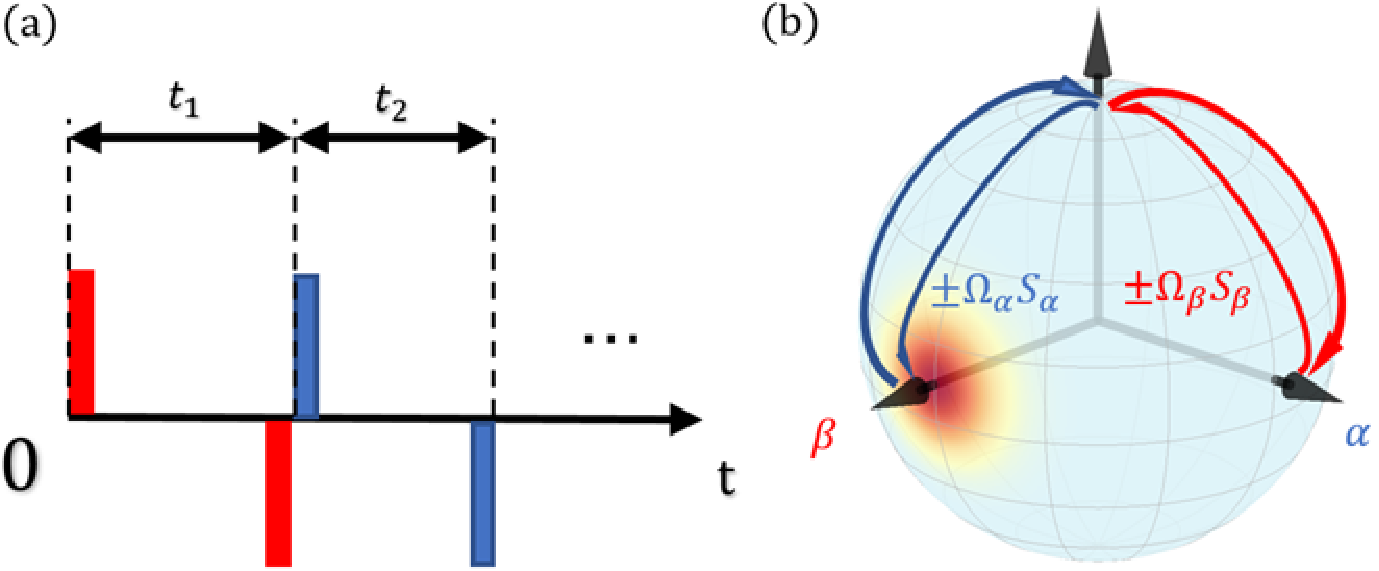}
\caption{(a) An illustration of the pulse sequences. The overall process
could be viewed as the repetition of (a). The red(blue) -up(down) pulse represents a $+\pi/2$($-\pi/2$) pulse along the $\alpha$($\beta$) axis. (b) The pulse method could also be
viewed as the cyclic rotation around the $\protect\alpha$ and $\protect\beta$ axis on the Bloch
sphere.}
\label{fig:6}
\end{figure}

To generate equivalent rime reversal $U_{2}=e^{iH_{\rm{LMG}}t}$, one way is to generate $H=-H_{\rm{LMG}}+kS^2$ since $[H, S^2]=0$, the total spin will not have an effect on the squeezing properties while $k$ is any constant. The whole time reversal process could be viewed as the repetitions of Fig.\ref{fig:6} (a), and each period is made up of the following:
a $-\pi/2$ pulse along the $\alpha$-axis, a free evolution for $t_1$, a $\pi/2$ along the $\alpha$-axis, a $-\pi/2$ along the $\beta$-axis, a free evolution for $t_2$, and a $\pi/2$ along the $\beta$-axis. The
period is $t_c=t_1+t_2$, neglecting the time needed for
applying the four $\pi/2$ pulses. Fig.\ref{fig:6}(b) shows that the
cyclic $\pm \pi/2$ could be viewed as rotations on the Bloch sphere. By
adjusting the relationship between $t_1$ and $t_2$, we can
transform the LMG interaction into equivalent time-reversed dynamics.

One potential strategy to generate time reversal is to apply multiple pulses along the $y$-axis and $z$-axis. For ${H_{\mathrm{LMG}}}=\chi({S_x^{2}+{\gamma}S_y^2})$, with the $y$-axis to be the $\alpha$-axis and $z$-axis to be the $\beta$-axis, the time evolution operates for a single period could be expressed as:
\begin{equation}
    \begin{split}
      {U_{yz}}&=R_{y,-\pi/2}e^{-i(S^2_x+\gamma S^2_y)\chi t_1}R_{y,\pi/2}\\
&R_{z,-\pi/2}e^{-i(S^2_x+\gamma
S^2_y)\chi t_2}R_{z,\pi/2} \\
&=e^{-i(S^2_{z}+ \gamma S^2_{y})\chi t_1}e^{-i(S^2_{y}+ \gamma S^2_{x})\chi
t_2}.
    \end{split}
\end{equation}
Using the Baker-Campbell-Hausdorff formula, we find $U_{yz}\approx e^{-i\chi
[S_{x}^{2}(\gamma t_2)+S_{y}^{2}(\gamma t_{1}+t_2)+S_{z}^2(t_1)]}$ for small $t_1$ and $t_2$ \cite{BCH}. In order to achieve equivalent time reverse, the relationship between $\gamma $ and $t_{2}/t_{1}$ should satisfy
\begin{equation}
\frac{t_{2}}{t_{1}}=\frac{1-2\gamma}{(1-\gamma)(1+\gamma)}.
\end{equation}%
Accordingly, we obtain the effective Hamiltonian
\begin{eqnarray}
H_{yz}^{\mathrm{eff}} =\frac{\chi_{\rm eff}}{\chi} H_{\rm LMG}=-\frac{\chi (\gamma^2-\gamma+1)}{-\gamma^2-2\gamma+2}(S_{x}^{2}+\gamma S_{y}^{2}),  
\end{eqnarray}
With the anisotropic parameter $\gamma$
ranging from 0 to 0.5, $\chi_{\rm eff}=-\frac{ (\gamma^2-\gamma+1)}{-\gamma^2-2\gamma+2}  < 0$, which means that we can generate equivalent time reverse by applying pulse sequences along the $y$-axis and $z$-axis.

\section{\label{sec:4}Noise analysis}

Now we analyze the performance of the time-reversal interferometry scheme under various kinds of noises, and here we focus on analyzing the impact of detection noise as well as the imperfect pulse. The former could be weakened by applying the time-reversal control, while the latter is introduced by the multiple pulse sequences involved in our scheme.

\subsection{\label{sec:level2}Detection noise}

The effect of detection noise can be described by the measuring operator $\hat{M}$: 
\begin{equation}
\hat{M} = \sum_{m,m'}  \Gamma_{m,m'} \left  | m '\right \rangle \left  \langle m\right |,
\end{equation}
where $\left  | m \right \rangle$ is the $m$th eigen state of ${S}_{z}$ and 
\begin{equation}
\Gamma_{m,m'} = \frac{e^{-(m-m')^2/(2\sigma)^2}}{\sum_{m'}e^{-(m-m')^2/(2\sigma)^2}},
\end{equation}
determined by parameter $\sigma$. This setup means if the system is in $\left  | m \right \rangle$ state, execute the measurement and there exists a probability of $\Gamma_{m,m'}$ to mistake $\left  | m \right \rangle$ for $\left  | m' \right \rangle$. $\Gamma_{m,m'}$ has a form of Gaussian convolution, which makes sense experimentally and mathematically. Now we define noise strength $\mathcal{N} = e^{-1/(2\sigma)^2}$,  $\hat{M}$ then could be expressed as:
\begin{equation}
\hat{M} = \frac{1}{1+ 2 \sum_{n} \mathcal{N}^{n^2}} \sum_{0\leq m\pm k,1\leq m\leq N+1}  \mathcal{N}^{k^2} \left  | m \pm k\right \rangle \left  \langle m\right |
\end{equation}
$\hat{M}$ could be regarded as a identity operator $\hat{I}$ attached by higher order term of $\mathcal{N}$, which would cause a correction term in the angular sensitivity $\Delta \phi$. Now we define the robustness coefficient $R$ as:
\begin{equation}
R = -\lg\frac{\partial \Delta \phi}{\partial \mathcal{N}},
\end{equation}
which reflects the response of measurement results to the detection noise. A larger R means less response, namely better robustness. When $\mathcal{N}$ is not too large (for example, $\mathcal{N}=0.1$, which means a probability of 0.2 to make an error just in the detection procedure), we can get the value of $R$ semi-analytically. For simplicity, we take $t_2 = t_1$. The detail is shown in Appendix~\ref{app:app3}.

\begin{figure}[tb]
\centering
\includegraphics[width = 0.45\textwidth]{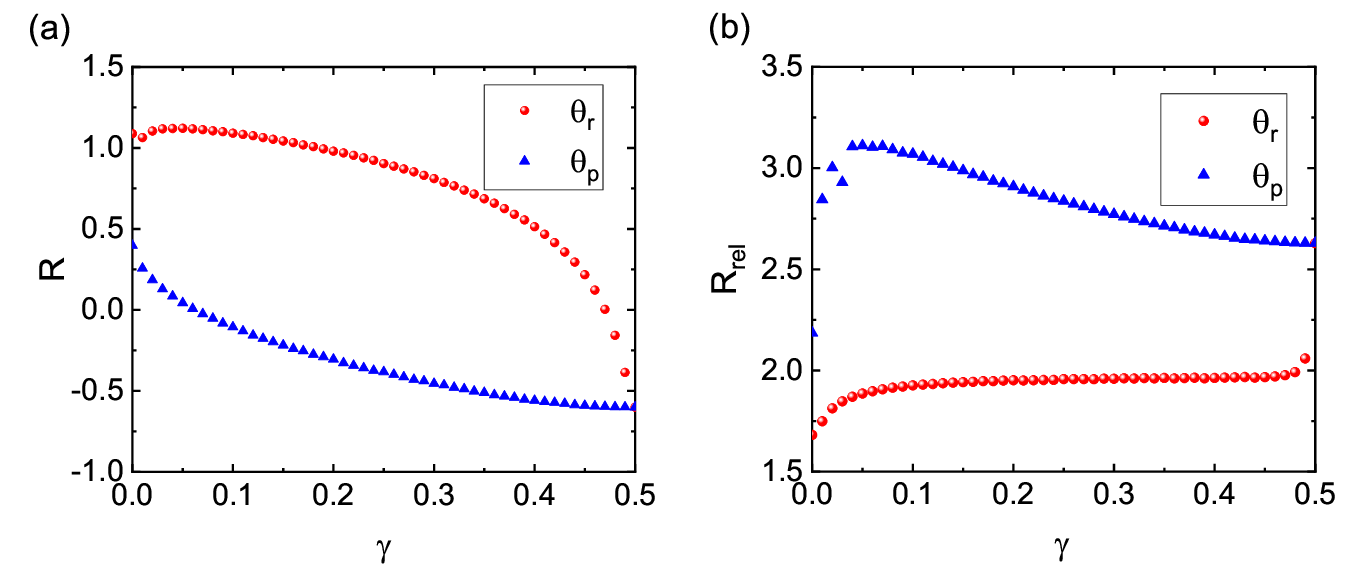}
\caption{(a)  Under the noise strength $\mathcal{N}=0.1$, the relationship between robustness coefficient $R$ and $\theta_{r}$, $\theta_{p}$ for different $\gamma$. (b) Under the noise strength $\mathcal{N}=0.1$, the relative robustness coefficient $R_{\rm rel}$ with $\theta_{r}$ and $\theta_{p}$ for different $\gamma$ compared to no time-reversed scheme.(N = 100). }
\label{fig:7}
\end{figure}

The conditions when ${S}_{\theta}$ is towards $\theta_{r}$ and $\theta_{p}$ are calculated respectively. The result is shown in Fig.~\ref{fig:7}(a). We can find that in most of the cases, the robustness is better when ${S}_{\theta}$ is towards $\theta_{r}$. More precisely, when ${S}_{\theta}$ is towards $\theta_{p}$, the additional fluctuation caused by detection noise is about 10 times as large as that of $\theta_{r}$, especially when $\gamma$ is small. When $\gamma$ gets larger, this gap is narrowing. When $\gamma = 0.5$, the two curves coincide, for $\theta_{r} = \theta_{p}$ then. 

In order to measure the noise suppression effect of our scheme, we also calculate the relative robustness coefficient: $R_{\rm rel} = R-R_0$, where $R_{0}$ is the robustness coefficient when $U_{2}$ is Identity (no time-reversed dynamic is applied). The result is shown in Fig.~\ref{fig:7}(b). It can be concluded that for both $\theta_{r}$ and $\theta_{p}$, the IBR method can reduce the effect of noise to $10^{-2}$ to $10^{-3}$, which reflects the superiority of our scheme in noise reduction. Conversely, the $R_{\rm rel}$ of $\theta_{p}$ performs better than that of $\theta_{r}$, and when $\gamma = 0.5$ it is the same as the two conditions.

\subsection{\label{sec:level2} 
 Imperfect pulse}

Apart from the detection noise, noise from imperfect pulses should also be taken into consideration. Firstly, our scheme assumes that the duration of each period is small enough to ignore the high-order term, which means the frequency for the pulse period is very high. Besides, the pulse area, pulse separation, and pulse phase are not always perfect. To verify the robustness of our scheme, we perform numerical simulations by adding Gaussian stochastic noises, i.e., assuming the fluctuating pulse areas, pulse separations and pulse phase are subject to Gaussian distribution with a standard deviation of different ranges of the average value. 

\begin{figure}[tb]
\centering
\includegraphics[width = 0.45\textwidth]{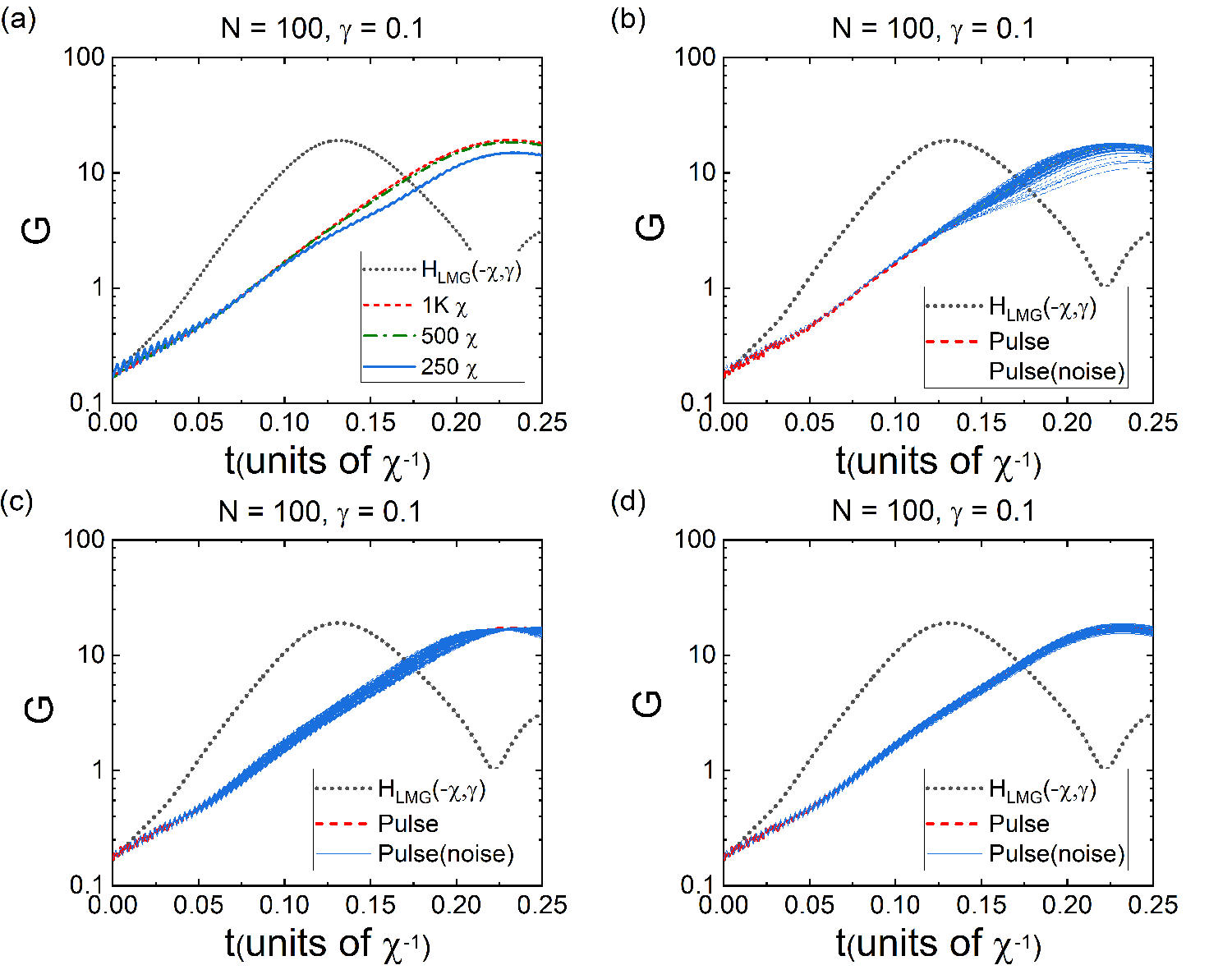}
\caption{ Numerical analysis of the influence of noises for our scheme with $N=100,\protect\gamma=0.1$. The blue curves crowding together denote the results of 100 independent simulations under different situations. (a) Evolution of the gain factor under different pulse frequencies. (b) Evolution
of the gain factor $G$ for 0.5\% level of Gaussian stochastic noise adding on the pulse area. (c) Evolution of the gain factor $G$ for 5\% level of Gaussian stochastic noise adding on the pulse separation. (d) Evolution of the gain factor $G$ for 0.1\% level of Gaussian stochastic noise adding on the phase of the pulse.  }
\label{fig:8}
\end{figure}

To quantify the influence of imperfect pulse, 
As shown in Fig.\ref{fig:8}, by adding the noise into the pulse parameters and repeating the simulations 100 times, we show that under the circumstances that the noise of for pulse frequency higher than 500$\chi$, noise in the pulse area less than 0.5$\%$, noise in pulse separation less than 5$\%$, and noise in pulse phase less than 0.1$\%$, our method can almost achieve the optimal performance under perfect time reverse, indicating the robustness of the pulse scheme, and set the bound for experimentalists to realize the dynamics.

As for the spin decoherence, our method will not introduce new resources of decoherence but expend the experimental time, and we consider the situation that the coherence time is long enough to ignore the impact of spin decoherence \cite{ibr2022lmg}, so we ignore the impact of extending the evolution time for spin decoherence for simplification.

\section{\label{sec:6}Summary}
In summary, we investigate the performance of nonlinear time-reversal interferometry with quadratic
collective-spin interaction. We design nonlinear interferometry based on that and optimize each part of this scheme with high precision and robustness with different anisotropic parameters. We find no tradeoffs in reaching the highest precision and robustness with the TAT interaction, which is an exceptional case in quadratic collective-spin interaction, and we prove it through numerical and analytic analysis. Besides, we find that OAT outperforms in robustness while TAT outperforms in precision. To achieve the time reversal in this interferometry, we propose a Floquet-driving method to generate the equivalent time reversal. With $H_{\rm LMG}=S_x^2+\gamma S_y^2$, we find that by making use of multi-pulse sequences along $y$-axis and $z$-axis, we could achieve effective time reverse. We also show that our scheme is robust to different kinds of noise, including detection noise, and imperfect pulses. Our work will significantly deepen the insight of making use of atomic systems to achieve high precision and high robustness in nonlinear interferometry and push the frontier of quantum metrology.

%%\bibliography{references}

\section{acknowledgement}

We thank Guoqing Wang and Qi Liu for the helpful discussions. This work is supported by the National Key R\&D Program of China (Grant No. 2023YFA1407600), and the National Natural Science Foundation of China (NSFC) (Grants No. 12275145, No. 92050110, No. 91736106, No. 11674390, and No. 91836302).

\appendix

\section{\label{app:app1}Mean-field approximation in perturbation encoding}

For the LMG model, we have
\begin{equation}
[{H},{S^2}] = 0,
\end{equation}
which means $S^2=S_x^2+S_y^2+S_z^2$ is constant during the evolution. Thus $H_{\mathrm{LMG}}(\chi, \gamma)$ is equivalent to:
\begin{equation}
H_{\mathrm{LMG}}(\chi, \gamma)-\chi \gamma S^2 = \chi[(1-\gamma)S_{x}^2-\gamma S_{z}^2].\label{appa2}
\end{equation}
When the unknown parameter is encoded through $S_{\theta}$, suppose $S_{\alpha}$  satisfies $S_{\alpha} = \langle {S}_{\alpha} \rangle + S_{\alpha 1}$, where $\alpha = x,z$ and $S_{\alpha 1}$ is the first order small quantity compared to $S_{\alpha}$. Applying this approximation to Eqs.~(\ref{appa2}), we can get the mean-field approximation of $H_{\mathrm{LMG}}(\chi, \gamma)$:
\begin{equation}
H_{\mathrm{LMG}}^{\mathrm{mf}}(\chi, \gamma) = 2\chi[(1-\gamma) \langle {S}_{x} \rangle S_{x}-\gamma \langle {S}_{z} \rangle S_{z}].
\end{equation}
For $\phi$ is far smaller than $S$, we can use S to represent $\langle {S}_{y} \rangle$. Consider $(\langle {S}_{x} \rangle, \langle {S}_{z} \rangle)$ as a point $(x,z)$ on the phase plane. Applying Heisenberg equation to $S_{x}$ and $S_{z}$, we have:
\begin{eqnarray}
\frac{dx}{dt}&=&\langle \frac{\partial {S}_{x}}{\partial t} \rangle \nonumber \\
&=&\langle \frac{1}{i\hbar} [{S}_{x}, {H}_{\mathrm{LMG}}^{\mathrm{mf}}]\rangle \nonumber\\
&=& 2\chi \gamma \langle {S}_{y} \rangle \langle {S}_{z} \rangle \nonumber\\
&=& 2\chi \gamma S z;\label{appa4}
\end{eqnarray}
\begin{eqnarray}
\frac{dz}{dt}&=&\langle \frac{\partial {S}_{z}}{\partial t} \rangle \nonumber \\
&=&\langle \frac{1}{i\hbar} [{S}_{z}, {H}_{\mathrm{LMG}}^{\mathrm{mf}}]\rangle \nonumber\\
&=& 2\chi (1-\gamma) \langle {S}_{y} \rangle \langle {S}_{x} \rangle \nonumber\\
&=& 2\chi (1-\gamma) S x,\label{appa5}
\end{eqnarray}
here we use the commutation relation of the collective spin operators:
\begin{equation}
[{S}_{\alpha},  {S}_{\beta}] = i\hbar \epsilon_{\alpha \beta \gamma} {S}_{\gamma},
\end{equation}
where $\alpha ,\beta ,\gamma$ denote the components in any three orthogonal directions, and $\epsilon_{\alpha \beta \gamma}$ is the Levi-Civita symbol. 
Combining Eqs.~(\ref{appa4}) and Eqs.~(\ref{appa5}), we get the evolution equation of x and z:
\begin{equation}
\begin{cases}
    \frac{d^{2}x}{dt^2} = (2\chi S)^{2} \gamma(1-\gamma)x\\
    \\
    \frac{d^{2}z}{dt^2} = (2\chi S)^{2} \gamma(1-\gamma)z
\end{cases},
\end{equation}
Solve the equations, we have
\begin{eqnarray}
x&=&Ae^{2\chi S \sqrt{\gamma(1-\gamma)} t}+Be^{-2\chi S \sqrt{\gamma(1-\gamma)} t} \label{appa8}\\
z&=& Ce^{2\chi S \sqrt{\gamma(1-\gamma)} t}+De^{-2\chi S \sqrt{\gamma(1-\gamma)} t},
\end{eqnarray}
where A, B, C, and D are undetermined coefficients. When $S_{\phi,\theta}$ has just added, we have:
\begin{equation}
\begin{cases}
    x_{0} =S \phi \sin\theta\\
    \\
    (\frac{dx}{dt})_{0} =2S\chi \gamma z_{0} = 2S^2\chi \gamma  \phi \cos\theta
\end{cases},
\end{equation}
apply which to Eqs.~(\ref{appa8}) we can determine the expression for x:
\begin{eqnarray}
x &= &\frac{S\phi}{2}[(\sin\theta+\sqrt{\frac{\gamma}{1-\gamma}}\cos\theta)
e^{2\chi S \sqrt{\gamma(1-\gamma)} t}\nonumber\\
&+&
(\sin\theta-\sqrt{\frac{\gamma}{1-\gamma}}\cos\theta)
e^{-2\chi S \sqrt{\gamma(1-\gamma)} t}].
\end{eqnarray}
Similarly, we can get the expression for z:
\begin{eqnarray}
z &= &\frac{S\phi}{2}[(\sqrt{\frac{1-\gamma}{\gamma}}\sin\theta+cos\theta)
e^{2\chi S \sqrt{\gamma(1-\gamma)} t}\nonumber\\
&-&
(\sqrt{\frac{1-\gamma}{\gamma}}\sin\theta-\cos\theta)
e^{-2\chi S \sqrt{\gamma(1-\gamma)} t}].
\end{eqnarray}
To ensure $\langle {S_{\phi}} \rangle$ reaches its maximum, we can equivalently ascertain a $\theta$ that makes the point $(x,z)$ on the phase plane as far away from the origin as possible during the evolution. The distance from $(x,z)$ to the origin is
\begin{eqnarray}
&&\sqrt{x^2+z^2} =\frac{S\phi}{2}e^{2\chi S \sqrt{\gamma(1-\gamma)} t}\times\nonumber\\
&&[(\sqrt{\gamma}+{\frac{1-\gamma}{\sqrt{\gamma}}})\sin\theta +
(\sqrt{1-\gamma}+{\frac{\gamma}{\sqrt{1-\gamma}}})\cos\theta]
,\nonumber\\\label{appa13}
\end{eqnarray}
where we omit the terms with $e^{-2\chi S \sqrt{\gamma(1-\gamma)} t}$. Eqs.~(\ref{appa13}) has maximum when 
\begin{equation}
(\sqrt{\gamma}+{\frac{1-\gamma}{\sqrt{\gamma}}})\sin\theta +
(\sqrt{1-\gamma}+{\frac{\gamma}{\sqrt{1-\gamma}}})\cos\theta
\end{equation}
reaches its maximum, which leads to our result:
\begin{equation}
\theta_{r} = \arcsin\sqrt{\gamma}.
\end{equation}
\section{\label{app:app3}Semi-analytical solution of robustness coefficient $R$}
When $\mathcal{N}$ is not too large, the main source of error is mistaking $\left  | m \right \rangle$ for $\left  | m\pm 1 \right \rangle$. The other term of $\hat{M}$ would vanish because of the higher order of $\mathcal{N}$. Thus $\hat{M}$ can be approximated by
\begin{equation}
\hat{M} \approx \frac{1}{1+ 2  \mathcal{N}}(\hat{M}_{0} + \mathcal{N}\hat{M}_{1}),\label{appb1}
\end{equation}
where $\hat{M}_{0}$ is simply the identity operator and $\hat{M}_{1} = \sum_{m\in \mathbb{N},m \pm 1 \ge 0}  \mathcal{N} \left  | m \pm 1\right \rangle \left  \langle m\right |$. In matrix form, $\hat{M}_{0}$ can be denoted as an identity matrix $I$, and $\hat{M}_{1}$ can be denoted as a matrix $M_1$ with non-zero element $M_{1}(k,k+1) = M_{1}(k,k-1) = 1$. 
Taking $M$ into consideration, the final state turns into: $M U_{2} e^{-i{S}_{\theta}\phi} U_{1} \left  | \hat{\boldsymbol {y}}\right \rangle$. Applying Eqs.~(\ref{appb1}) into Eqs.~(\ref{15}), we can get the expression of $ \langle S_{m}^{\phi} \rangle$ under the  perturbation of $\hat{M}_{1}$: 
\begin{align}
    \langle S_{m}^{\phi} \rangle & = \left  \langle \hat{\boldsymbol {y}}\right | U_{1}^{\dagger} U_{\phi,\theta}^{\dagger} U_{2}^{\dagger} M^{\dagger}{S}_{m} M U_{2} U_{\phi,\theta} U_{1} \left  | \hat{\boldsymbol {y}}\right \rangle \nonumber \\
    & \approx  \frac{1}{(1+2\mathcal{N})^2}\Big\{\left  \langle \hat{\boldsymbol {y}}\right | U_{1}^{\dagger} e^{i{S}_{\theta}\phi} U_{2}^{\dagger} {S}_{m} U_{2} e^{-i{S}_{\theta}\phi} U_{1} \left  | \hat{\boldsymbol {y}}\right \rangle \nonumber \\
    & +  \mathcal{N} \left  \langle \hat{\boldsymbol {y}}\right | U_{1}^{\dagger} e^{i{S}_{\theta}\phi} U_{2}^{\dagger} \{M_{1},{S}_{m}\} U_{2} e^{-i{S}_{\theta}\phi} U_{1} \left  | \hat{\boldsymbol {y}}\right \rangle \Big\} \nonumber\\
    & = \frac{i\phi}{(1+2\mathcal{N})^2} \Big\{ \left  \langle \hat{\boldsymbol {y}}\right | [{S}_{\theta}(U_{1}),{S}_{m}(U_{2}U_{1})] \left  | \hat{\boldsymbol {y}}\right \rangle \nonumber \\
    & +  \mathcal{N} \left  \langle \hat{\boldsymbol {y}}\right | [{S}_{\theta}(U_{1}),\{M_{1},{S}_{m}\}(U_{2}U_{1})] \left  | \hat{\boldsymbol {y}}\right \rangle \Big\}\nonumber\\
    &+ O(\phi^2),\label{appb2}
\end{align}
where we define $\{M_{1}, {S}_{m} \} = M_{1} {S}_{m}+ {S}_{m}M_{1}$. Expanding to the first order of $\mathcal{N}$, then we can get an approximation of $\partial_{\phi}  \langle S_{m}^{\phi} \rangle$:
\begin{align}
    \partial_{\phi}  \langle S_{m}^{\phi} \rangle & = \frac{i}{(1+2\mathcal{N})^2} \Big\{ \left  \langle \hat{\boldsymbol {y}}\right | [{S}_{\theta}(U_{1}),{S}_{m}(U_{2}U_{1})] \left  | \hat{\boldsymbol {y}}\right \rangle \nonumber \\
    & +  \mathcal{N} \left  \langle \hat{\boldsymbol {y}}\right | [{S}_{\theta}(U_{1}),\{M_{1},{S}_{m}\}(U_{2}U_{1})] \left  | \hat{\boldsymbol {y}}\right \rangle \Big\}\nonumber\\
    & \approx i \Big\{ (1-4\mathcal{N})\left  \langle \hat{\boldsymbol {y}}\right | [{S}_{\theta}(U_{1}),{S}_{m}(U_{2}U_{1})]  \left  | \hat{\boldsymbol {y}}\right \rangle 
    \nonumber \\
    & +  \mathcal{N} \left  \langle \hat{\boldsymbol {y}}\right | [{S}_{\theta}(U_{1}),\{M_{1},{S}_{m}\}(U_{2}U_{1})] \left  | \hat{\boldsymbol {y}}\right \rangle \Big\}\nonumber\\
    &+ O(\phi^2).\label{appb3}
\end{align}
Applying Eqs.~(\ref{appb3}) to Eqs.~(\ref{14}) and expanding to the first order of $\mathcal{N}$, we can get the corrected $\Delta \phi$:
\begin{equation}
\frac{\partial \Delta \phi}{\partial \mathcal{N}} = \left[\frac{\Delta S_{m}^{\phi}}{\partial_{\phi} \langle {S}_{m}^{\phi} \rangle } \Big\{4- \frac{\left  \langle \hat{\boldsymbol {y}}\right | [{S}_{\theta}(U_{1}),\{M_{1},{S}_{m}\}(U_{2}U_{1})] \left  | \hat{\boldsymbol {y}}\right \rangle}{\left  \langle \hat{\boldsymbol {y}}\right | [{S}_{\theta}(U_{1}),{S}_{m}(U_{2}U_{1})]  \left  | \hat{\boldsymbol {y}}\right \rangle }
     \Big\} \right] _{\phi = 0},
\end{equation}
and go further to figure out R.

\bibliography{main}

\begin{thebibliography}{64}
\bibitem [{TAT()}]{TAT}%
  \BibitemOpen
  \href@noop {} {}\bibinfo {howpublished} {One general Hamiltonian for the TAT
  interaction is $H_{\rm TAT}=S_xS_y+S_yS_x$ for $θ=\pi/2$ $\phi=\pm \pi/4$ .
  By changing the initial states and twisting axes, the TAT interaction could
  also be expressed as $H_{\rm TAT}=S_x^2-S_z^2$. While the constant of
  $S^2=S_x^2+S_y^2+ S_z^2$ will not influence the properties of spin squeezing,
  we simply ignore it, and that makes the TAT interaction as $H_{\rm
  TAT}=0.5(S_x^2-S_z^2)$.}\BibitemShut {Stop}%
\bibitem [{BCH()}]{BCH}%
  \BibitemOpen
  \href@noop {} {}\bibinfo {howpublished} {The BCH formula has the expression
  as $e^A e^B = {\rm
  exp}({A+B+\frac{1}{2!}[A,B]+\frac{1}{3!}(\frac{1}{2!}[A,[A,B]+\frac{1}{2!}[[A,B],B)+...}).
  $.}\BibitemShut {Stop}%
\end{thebibliography}

\end{document}